\documentstyle[tighten,preprint,prb,aps]{revtex}
\begin{document}
\draft
\preprint{}
\title{Magnetoconductivity of quantum wires\\ with elastic
and inelastic scattering}
\author{Henrik Bruus}
\address{Applied Physics, Yale University, \\P.O.Box 2157,
   New Haven, Connecticut 06520, USA}
\author{Karsten Flensberg}
\address{Nordita,
Blegdamsvej 17, Copenhagen 2100 \O, Denmark}
\author{Henrik Smith}
\address{Niels Bohr Institute, \O rsted Laboratory,\\
Universitetsparken 5, Copenhagen 2100 \O, Denmark}
\date{March 15, 1993}
\maketitle
\begin{abstract}
We use a Boltzmann equation to determine  the magnetoconductivity
of quantum wires. The presence of
a confining potential in addition to the magnetic field
removes the degeneracy of the Landau levels and
allows one to associate a group velocity
with each single-particle state. The distribution function
describing the occupation of these single-particle states
satisfies a Boltzmann equation, which may be solved
exactly in the case of impurity scattering.
In the case where the electrons scatter against
both phonons and impurities we solve numerically - and
in certain limits analytically - the integral
 equation for the distribution function, and determine
the conductivity as a function of temperature and magnetic field.
The magnetoconductivity
exhibits a maximum at a tempe\-rature, which depends on the relative
strength of the  impurity and electron-phonon scattering, and shows
oscillations when the Fermi energy or the magnetic field is varied.
\end{abstract}

\narrowtext

\section{Introduction}
The possibility of making quasi-one-dimensional quantum
wires,  due to   advances in microfabrication technology,
has greatly stimulated the interest in the
 transport properties of such low-dimensional systems.
A number of recent  papers
have treated the magnetoconductivity of quantum wires \cite{beenakker}
using a variety of different theoretical approaches.
The suppression of scattering between edge-states,
due to the presence of the magnetic field, was discussed
by several authors \cite{martin,maslov,komi}. The effects of
impurity and boundary scattering
on the Hall effect in quantum wires were treated
by Akera and Ando \cite{ando}, starting from a Boltzmann equation. The
Kubo  method \cite{Kubo} has been used   for considering
the effect of the coupling to optical phonons \cite{vasilo,mori}
and the influence of impurity scattering \cite{bruus} on the
magnetotransport through quantum wires.
          Momentum- and energy-balance equations have been derived
\cite{tso,boiko} for  a quasi-one-dimensional gas of electrons
in a magnetic field. Other authors \cite{wagner} have used
the Keldysh method to discuss
magnetotransport in the presence of impurity
scattering, while the formalism due to Landauer \cite{Landauer}
and B\"{u}ttiker \cite{buttiker}
was  employed in Ref.~\onlinecite{shepard}
to treat the influence of disorder
on the  Hall effect in quantum wires.

Transport in strong magnetic fields has traditionally
been treated within the Kubo-formalism, since
this allows one to take fully into account the quantization
of the motion of an electron in a magnetic field. The
resulting diagrammatic expansion of the conductivity
requires the consideration of both self-energy
effects and vertex corrections. It was
demonstrated in Ref.~\onlinecite{bruus} that it is essential
to include vertex corrections in order to
determine the effect of impurity scattering on the magnetoconductivity
of a quantum wire. The physical reason for this
is that only back-scattering from the impurities  contributes to the
resistivity.

In the present paper we
 use a Boltzmann equation to determine  the transport properties
of quantum wires formed by  additional
confinement of the two-dimensional electron gas of
{\it e.g.\/}\ a GaAs-GaAlAs
heterojunction. The presence of
a confining potential in addition to the magnetic field
removes the degeneracy of the Landau levels and
allows one to associate a group velocity
with each single-particle state. The distribution function
describing the occupation of these single-particle states
satisfies a Boltzmann equation, which we solve
exactly in the case of impurity scattering. The resulting
magnetoconductivity agrees with that obtained in
Ref.~\onlinecite{bruus}
within the Kubo-approach. In the case of electron-phonon scattering we
 solve numerically - and in certain limits analytically -
 the integral equation for the distribution function.
The conductivity is determined as a function of
temperature and magnetic field in the case when both
impurities and phonons contribute to the scattering
of the electrons.
 Our paper extends  previous work
discussed above by giving a more complete account of the
simultaneous influence of electron-impurity
and electron-phonon scattering. By starting from the Boltzmann equation
we include vertex corrections from the outset, thus
simplifying the formal development  in comparison to
that involved in the use of the Kubo formalism.

Apart from its simplicity,
the use of the Boltzmann equation has the additional advantage
that it allows one to go beyond linear response
theory in a straightforward manner.
Although the present paper treats in detail
only the regime of linear response,
our approach may readily be  generalized to take nonlinear effects
into account.

Since conduction in a quantum wire involves current transport
in one dimension, it is necessary to consider  the role
of localization. This was done in Ref.~\onlinecite{suhrke}, where it
was shown that weak localization in quantum wires
is destroyed in a magnetic field greater than a critical value
 $B_c$, where the critical magnetic field $B_c$
 apart from numerical factors is given by
$B_c\simeq h\ell/eL_{\phi}W^2$. Here $\ell$ is the elastic scattering
length, $W$ the width of the quantum wire,
 while $L_{\phi}$ is the phase coherence length. For
the parameters considered in Ref.~\onlinecite{suhrke} this yields
a $B_c$ somewhat less than $0.1$ T.
In what follows we shall assume that the magnetic field is
 sufficiently strong that localization effects are
always negligible. The transport properties of the quantum
wires may then be obtained from a Boltzmann equation,
in which the quantizing effect of the magnetic field is
incorporated  in the manner  described below.
We shall also disregard
Coulomb-blockade effects which have been shown to be of importance
in some systems \cite{staring}.

The paper is organized as follows.
In section~2  we introduce the single-particle states derived from the
presence of a confining potential in addition to a
homogeneous magnetic field. The Boltzmann equation and its
solution is discussed in section~3. In section~4
we treat scattering due to impurities, while  scattering from
impurities and acoustical  phonons forms the subject of section~5,
where we consider both deformation-potential and
piezoelectric coupling.
Finally section~6 deals with the case where
scattering from  impurities and optical phonons
is important.

\section{The single-particle energy spectrum}
We consider an electron moving in the $xy$-plane under the influence
of a constant magnetic field in the $z$-direction. In addition
a parabolic confinement potential
limits its motion  in the $y$-direction whereby a wire is formed in
the $x$-direction.
The Hamiltonian for such an electron is
\begin{equation}
\hat{H}=\frac{1}{2m^*}(\bbox{p}+e\bbox{A})^2+\frac{1}{2}Ky^2,
\end{equation} $m^*$ being the effective (band)
mass of the electron. For simplicity we have neglected the Zeeman
splitting, since we shall be mainly  concerned with quantum wires
in GaAs-based structures,
where the Zeeman energy is a few percent of the cyclotron energy,
$\hbar \omega_c$, to be defined below.  The electron spin therefore
only enters as a factor of 2 in
 the expression for the current density.
In the $x$-direction we impose periodic boundary conditions, and
for the vector-potential $\bbox{A}$
we use the Landau-gauge, $\bbox{A} = B(-y,0,0)$.
 Consequently
the eigenfunctions of the Hamiltonian are products of plane waves
and harmonic-oscillator functions involving
Hermite polynomials ${\rm H}_n$,
\begin{eqnarray}
\Psi_{nk}(x,y) &=& \frac{1}{\sqrt{L}} e^{ikx}
\frac{1}{\sqrt{\sqrt{\pi} 2^n n! \ell_h}} \nonumber \\
&&\times{\rm H}_n\left(\frac{y-y_k}{\ell_h}\right)
\exp\left(-\frac{(y-y_k)^2}{2\ell_h^2}\right),
\label{eigen}
\end{eqnarray}
with $L$ being the length of the quantum wire.
The center coordinate, $y_k$, and the characteristic length, $\ell_h$,
of the harmonic oscillator is specified below. The wave function in
the $z$-direction is taken to be a delta function.

In the case where $K$ is zero and only the magnetic field is present
the natural system of units is based on the cyclotron frequency
$\omega_c$ and the magnetic length $\ell_c$, given by
\begin{equation}
\omega_c=\frac{eB}{m^*}, \hspace{10mm} \ell_c^2=\frac{\hbar}{eB}.
\end{equation}
When $K$ is different from zero, we define the
quantity $\gamma$, which describes the relative strength of the
confinement potential,  by
\begin{equation}
\gamma \equiv \frac{K}{m^*\omega_c^2},
\end{equation} and the natural units are now based on the hybrid
frequency, $\omega_h$, and the hybrid
characteristic length, $\ell_h$, given by
\begin{equation}
\omega_h = (1+\gamma)^{ \frac{1}{2}} \omega_c, \hspace{10mm}
\ell_h = (1+\gamma)^{-\frac{1}{4}} \ell_c.
\end{equation}

By inserting the function (\ref{eigen}) in the
Schr\"{o}dinger-equation
\begin{equation}
\hat{H}\Psi_{nk} =\varepsilon_{nk}\Psi_{nk},
\label{sch}
\end{equation}
one finds that (\ref{sch}) is satisfied, provided
\begin{equation}
\varepsilon_{nk}=\hbar\omega_h \left(n+\frac{1}{2}\right)
+\frac{\hbar^2k^2}{2m},\;\;n=0,1,2,\ldots,
\label{energy}
\end{equation}
where $m$ is a renormalized mass defined by
\begin{equation}
m=\frac{1+\gamma}{\gamma}m^*.
\end{equation}
The center of the harmonic-oscillator wavefunctions is given by
\begin{equation}
y_k= \frac{\omega_c\hbar k}{m^*\omega_h^2} =
(1+\gamma)^{-\frac{1}{2}} k\ell_h^2.
\end{equation}

The confinement potential $Ky^2/2$  removes the de\-gene\-racy
of the Landau-levels through the $k$-de\-pen\-dence ex\-hibited in
(\ref{energy}). Note that the dispersion relation  is that of a
free particle with a renormalized mass $m$.
When the strength of the confinement potential goes to zero,
the renormalized mass $m$ becomes infinite (corresponding to
degenerate Landau-levels), while in the opposite limit
($\gamma\rightarrow \infty$), the renormalized mass
becomes equal to the effective mass $m^*$.
The $k$-dependent energy levels are plotted in Fig.\ 1, where
we have also indicated the position of the Fermi level corresponding
to a definite electron density.

In the following we shall describe
conduction  through the quantum wire
on the basis of the Boltzmann equation for the distribution
function of the excitations with energy $\varepsilon_{nk}$.
The group velocity $v$ of the excitations is seen to be
\begin{equation}
v_{nk}=\frac{1}{\hbar}
\frac{\partial\varepsilon_{nk}}{\partial k}=\frac{\hbar k}{m}.
\label{vk}
\end{equation}
If the confinement potential is not simply parabolic, the
relation between $v_{nk}$ and $k$ becomes more complicated
than Eq.\ (\ref{vk}), but
the method described in the following is still applicable,
provided one modifies  the
 group velocity accordingly, and incorporates the new wavefunctions
into the matrix elements appearing in the collision probability.

\section{The Boltzmann equation}
We consider the distribution function $f_{nk}$ for
the excitations specified by the energy $\varepsilon_{nk}$,
with group velocity $v_{nk}=\hbar k/m$.
Unlike the usual semi-classical description of transport
in a magnetic field, the magnetic field has here been taken
into account  from the very beginning, in defining the excitations
with dispersion relation (\ref{energy}).
The effect of
an electric field $E$ in the $x$-direction is, however,
 included in the usual manner through the acceleration equation
\begin{equation}
\hbar \dot{k}=-eE.
\label{acc}
\end{equation}

The Boltzmann equation \cite{transport} is
\begin{equation}
\frac{\partial f_{nk}}
{\partial t}+\dot{k}\frac{\partial f_{nk}}{\partial k}
=\left(\frac{\partial f_{nk}}{\partial t}\right)_{\rm coll},
\end{equation}
where the right hand side of the equation contains the
 collision term.
Since we are interested in the linear
response to a static electric field,
the time-derivative $\partial f/\partial t$ is equal to
zero, while $f_{nk}$ in the second term on the left
hand side may be
replaced by the equilibrium function $f^0_{nk}$.
Using Eq.\ (\ref{acc}) the Boltzmann equation
then becomes an inhomogeneous
integral equation of the form
\begin{equation}
\frac{eE}{k_{\rm B}T}v_{nk}f^0_{nk}(1-    f^0_{nk})
=\left(\frac{\partial f_{nk}}{\partial t}\right)_{\rm coll},
\end{equation}
with the integral term on the right hand side
 - the collision term - to be specified below.

Before we consider specific scattering mechanisms, let us
write down the conductivity $\sigma$ in terms of the
distribution function.
In the present work we restrict ourselves to the calculation
of the diagonal conductivity
element $\sigma_{xx}\equiv \sigma$.
The current density $j$ is
\begin{equation}
j=\sigma E
=-2e\sum_{n}^{}\int_{-\infty}^{\infty}\frac{dk}{2\pi}v_{nk}f_{nk}.
\end{equation}
We shall introduce the deviation function $\psi_{nk}$
by the  definition
\begin{equation}
f_{nk}  \equiv   f^0_{nk}  +   f^0_{nk}(1- f^0_{nk})\psi_{nk}.
\end{equation}
Then the current density becomes
\begin{equation}
j=-2e\sum_{n}^{}\int_{-\infty}^{\infty}\frac{dk}{2\pi}v_{nk}
  f^0_{nk}(1- f^0_{nk})\psi_{nk}.
\end{equation}
Since we shall be dealing with distribution functions
that change sign upon $k \rightarrow -k$, we may restrict $k$
to positive values and work instead with the
distribution function as a function of energy:
$\psi_n(\varepsilon)$.
In this notation $k$ is then a function of energy, and
we denote the $k$-value which solves
$\varepsilon=\varepsilon_{nk}$ by
$k_n(\varepsilon)$. Furthermore, by introducing the function
$\phi_n$ defined by
\begin{equation} \label{phi}
\phi_n \equiv \frac{k_{\rm B}T}{eE}\psi_n,
\end{equation}
we may express the conductivity as
\begin{equation}\label{cond}
\sigma = \frac{2e^2}{h} 2\sum_n \int_0^{\infty} d\varepsilon\,
\left(-\frac{\partial f^0(\varepsilon)}{\partial \varepsilon}\right)
\phi_n(\varepsilon)
\end{equation}
since  $ dk  =  d\varepsilon /\hbar v $.
Note that $\phi$ has dimension of  length.
The chemical potential entering the equilibrium distribution
function $f^0$ will be assumed to be independent of temperature,
equal to $\varepsilon_{\rm F}$, since the
quantum wire under typical experimental conditions is in
 contact with a large reservoir of electrons.

The Boltzmann equation considered in this
paper is  an inhomogeneous integral equation of the form
\begin{equation}
X_{nk} = H \psi_{nk},
\end{equation}
where
\begin{equation}
X_{nk}=-\frac{eE}{k_{\rm B}T}v_{nk}f^0_{nk}(1-    f^0_{nk}),
\end{equation}
while the integral operator $H$ is defined by
\begin{equation}
\left(\frac{\partial f_{nk}}{\partial t}\right)_{\rm coll} =
-H \psi_{nk}
\end{equation}
with
\begin{equation}
H \psi_{nk} =
     \sum_{n'}^{}\int_{-\infty}^{\infty}\frac{L}{2\pi}dk'
K_{nn'}(k,k') (\psi_{nk}-\psi_{n'k'}).
\label{intop}
\end{equation}
Here $K$ is an integral kernel to be specified in later sections.

By introducing the scalar product $(A,B)$ through the definition
\begin{equation}
(A,B)=\sum_{n}^{}\int_{-\infty}^{\infty}\frac{dk}{2\pi}A_{nk}B_{nk}
\end{equation}
the conductivity may be written as
\begin{equation}
\sigma =\frac{2k_{\rm B}T}{E^2}(X,\psi).
\end{equation}
In cases where we cannot solve explicitly for $\psi$, it
is useful to employ a variational principle, which yields
a lower  bound on $\sigma$ by virtue of the Schwarz inequality
\begin{equation}
(\psi,H\psi)(U,HU)\geq (U, H\psi)^2,
\end{equation}
where $U$ is an arbitrary trial function.
Since $H\psi=X$, this gives a lower bound on the conductivity
\begin{equation}
\sigma\geq\frac{2k_{\rm B}T}{E^2}\frac{(X,U)^2}{(U,HU)}.
\label{bound}
\end{equation}
This lower bound will be used to determine the
low-temperature conductivity when the scattering is due to
acoustic phonons, both with and without impurities,
and the resulting analytic expressions are compared to the results
of the numerical solution of the integral equation.

We shall always  write our calculated values of
 the conductivity, as limited by the various scattering mechanisms
to be considered in subsequent sections,  in the form
\begin{equation}
\sigma=\frac{2e^2}{h}l,
\label{form}
\end{equation}
where $l$ has the dimension of
a length. As may be expected, in the presence of impurity scattering
$l$ tends to a finite value as $T$ tends to zero, while
it increases exponentially in the absence of impurity scattering, when
the electrons are  scattered only by phonons.

\section{Impurity scattering}
In this section we consider the case of elastic impurity scattering
treated recently in Ref.~\onlinecite{bruus}. Our aim is
 to show how the results of Ref.~\onlinecite{bruus},
which were obtained by use of
 the Kubo-formalism, are derived within the present framework.
The collision integral is
\begin{equation}
\left(\frac{\partial f_{nk}}{\partial t}\right)_{\rm coll}=
-\sum_{n'}^{}\int_{-\infty}^{\infty}\frac{L}{2\pi} dk'
w_{nn'}(k,k') (f_{nk} -f_{n'k'}),
\end{equation}
since the scattering is elastic. According to the Golden Rule we
have
\begin{equation}
w_{nn'}(k,k')=\frac{2\pi}{\hbar}|\langle nk|V|n'k'\rangle|^2
\delta (\varepsilon_{nk}-\varepsilon_{n'k'}).
\end{equation}
The square of the matrix-element $\langle nk|V|n'k'\rangle$
for scattering from impurities may be
written as
\begin{eqnarray}
|\langle nk|V|n'k'\rangle|^2 &=&
\int d^2\bbox{r}\int d^2\bbox{r'}V(\bbox{r})  V(\bbox{r'}) \nonumber \\
&& \times\langle nk|\bbox{r}\rangle
\langle \bbox{r}|n'k'\rangle
\langle n'k'| \bbox{r'}\rangle
\langle \bbox{r}'|nk\rangle.
\end{eqnarray}

We shall  perform an ensemble average
over the distribution of impurities, corresponding to
the replacement
\begin{equation}
V(\bbox{r})
V(\bbox{r}') \; \rightarrow \; \langle V(\bbox{r})
V(\bbox{r}') \rangle = F(\bbox{r, r'}).
\end{equation}
When the impurity potential may be
approximated by a delta-func\-tion in space,
 the function $ F$ is proportional to the
 delta-function $\delta(\bbox{r-r'})$, corresponding to a
``white-noise'' model
\begin{equation}
F(\bbox{r,r'})=A^2n_{\rm imp}\delta(\bbox{r-r'}).
\end{equation}
Here we have introduced the number of impurities
per unit area $n_{\rm imp}$
together with the constant $A$, which denotes
the magnitude of the matrix-element for scattering from a single
impurity. The dimension of $A$ is that of an energy times an area.

In order to distinguish between states of equal energy but
opposite sign of $k$, we introduce the ``branch-index'' $s =\pm 1 $.
\begin{equation}
k_{ns}(\varepsilon) \equiv s k_n(\varepsilon)=
s\sqrt{\frac{2m}{\hbar^2}(\varepsilon-(n+\frac{1}{2})\hbar\omega_h)},
\label{kf}
\end{equation}
provided
\begin{equation}
(n+\frac{1}{2})\hbar\omega_h\leq\varepsilon.
\label{ineq}
\end{equation}
The density of states at the Fermi energy $\varepsilon_{\rm F}$
is  inversely proportional to
the $s$-independent Fermi velocity
$v_n(\varepsilon_{\rm F}) =\frac{\hbar}{m} k_{n}(\varepsilon)$.
Finally, since $\phi_{ns}$ is odd in $k$ and hence
changes sign with $s$, we may write  $\phi_{ns}=s\phi_{n}$. The
problem then becomes that of solving the matrix equation
\begin{equation}
s=\sum_{n'\;s'}B_{ns}^{n's'}(s\phi_{n}-s'\phi_{n'}),
\label{matr}
\end{equation}
and calculating the conductivity according to Eq.\ (\ref{cond}).
The matrix $B_{ns}^{n's'}$ is positive
and symmetric with respect to interchange of $ns$ with $n's'$.
Its elements are
\begin{equation}
B_{ns}^{n's'}=
\frac{1}{\hbar^2}
A^2n_{\rm imp}\frac{1}{v_nv_{n'}}\frac{1}{\ell_h\sqrt{2\pi}}
F_{ns}^{n's'}.
\label{ele}
\end{equation}
Here we have defined  the following dimensionless quantities
\begin{equation}
F_{ns}^{n's'}= \sqrt{2\pi}\;\ell_h\int_{-\infty}^{\infty}
dy|u_n(y-y_{k_{ns}})|^2|u_{n'}(y-y_{k_{n's'}})|^2.
\end{equation}

In Fig.\ 2 we show the result of solving the matrix equation
(\ref{matr})
and calculating the conductivity according to Eq.\ (\ref{cond})
as a function of the Fermi energy $\varepsilon_{\rm F}$,
for different choices of the
parameter $\gamma$ determining the strength of
the confining potential. The pronounced structure
is due to the opening of  new ``channels'' for scattering
each time the Fermi energy
exceeds $(n+1/2)\hbar\omega_h$. At finite temperature this
structure is somewhat smeared, as evidenced by the figure,
but still clearly visible.

When  $\hbar\omega_h/2\leq
\varepsilon_{\rm F}\leq 3\hbar\omega_h/2$
only the lowest Landau-level is involved,
corresponding to $n=n'=0$ in Eq.\ (\ref{matr}). Then the solution
becomes particularly simple, $\phi_0=1/2B_{0+}^{0-}$.
By means
of the Sommerfeld-expansion we obtain the following low-temperature
expression for the conductivity $\sigma={2e^2}/{hB_{0+}^{0-}}$,
\begin{eqnarray}
\sigma&=&\frac{2e^2}{h}l_h\sqrt{2\pi}\frac{\hbar^2}
{A^2n_{\rm imp}}\frac{\hbar^2k_{\rm F}^2}{m^2}
\exp \left(2k_{\rm F}^2l_h^2/[1+\gamma]\right) \nonumber \\
&&
\times \left[1+\frac{8\pi^2}{3\gamma^2}\left(1
+\frac{1+\gamma}{k_{\rm F}^2l_h^2}\right)
\left(\frac{k_{\rm B}T}{\hbar\omega_h}\right)^2\right],
\label{cond1}
\end{eqnarray}
where $k_{\rm F}=k_{0+}(\varepsilon_{\rm F})$ is the Fermi momentum.
The conductivity increases with temperature because  electrons
are excited into states which
have a smaller overlap with the states
at the opposite branch. Consequently the electrons
have a smaller back-scattering
probability. The result (\ref{cond1}) is
valid only to lowest order in $(k_{\rm B}T/\hbar\omega_h)^2$. In
obtaining our numerical results we do not use the
Sommerfeld-expansion, but calculate directly the integral over energy
according to the  expression (\ref{cond}).

As an illustration of the matrix-inversion involved in the
solution of the  Boltzmann equation
we shall  also give the expression for the zero-temperature
conductivity in the case when  $3\hbar\omega_h/2\leq
\varepsilon_{\rm F}\leq 5\hbar\omega_h/2$. By utilizing the
symmetry $\phi_{n+}=-\phi_{n-}$ the $4\times 4$ matrix-equation
becomes a  $2\times 2$ equation which may readily be solved
for $\phi_{0+}$ and $\phi_{1+}$.
The resulting conductivity $\sigma =(2e^2/h)2(\phi_{0+}+\phi_{1+})$ is
 at $T=0$ K given by
\widetext
\begin{equation}
\sigma =\frac{2e^2}{h}\frac{2( 2B_{0+}^{1+}+  B_{1+}^{1-}+
B_{0+}^{0-})}{2(B_{0+}^{1+}
B_{0+}^{1-} + B_{0+}^{0-}B_{1+}^{1-}) +
( B_{0+}^{1+}+  B_{0+}^{1-})(  B_{0+}^{0-}+B_{1+}^{1-}  )}.
\label{cond0}
\end{equation}
\narrowtext
 The result (\ref{cond0}) yields the conductivity
at zero temperature, when the Fermi energy $\varepsilon_{\rm F}$
satisfies the condition $3\hbar\omega_h/2\leq\varepsilon_{\rm F}\leq
5\hbar\omega_h/2$,
in agreement
with Ref.~\onlinecite{bruus}.
The generalization to more Landau-levels is straightforward. The
$B$-matrix in Eq.\ (\ref{matr}) is in general a
$2(n_0+1)\times 2(n_0+1)$
matrix given by Eq.\ (\ref{ele}),
where $n_0$ is the quantum number associated
with the highest occupied Landau-level at $T=0$ K.

It is convenient to relate the conductivity calculated in the
following  two sections to the
zero-temperature conductivity $\sigma_{2D}$
associated with motion in two
dimensions in the absence of a magnetic field and
a confinement potential. The latter is given by
\begin{equation}
\sigma_{2D} = \frac{e^2 n_{2D}}{m^*}{\tau_{\rm imp}}
=en_{2D}\mu_{\rm imp},
\end{equation}
where $n_{2D}$ is the electron sheet density and
$\mu_{\rm imp}$ is
the zero-field, zero-temperature mobility,
\begin{equation}
\mu_{\rm imp} =  \frac{e\tau_{\rm imp}}{m^*}
\end{equation}
with
\begin{equation}
 \frac{1}{\tau_{\rm imp}} =
\frac{m^* A^2 n_{\rm imp}}{\hbar^3}.
\end{equation}
We shall thus express our calculated
conductivity in units of $(2e^2/h)l_{\rm imp}$, where
\begin{equation}
l_{\rm imp}= \tau_{\rm imp}\omega_cl_c
\end{equation}
is a characteristic length which depends on the impurity content
as well as the magnetic field.

\section{Electron-phonon scattering}
Next we turn to the consideration of inelastic processes.
The present section treats the case where the electrons are
scattered by
acoustical phonons due to the combined  deformation-potential
and piezoelectric coupling, while the following section
discusses the
coupling of electrons to optical phonons.

Before specializing to a particular model let us write down the
contribution to the collision operator from the scattering against
phonons. The phonons are assumed to be in thermal equilibrium.
Then the integral operator $H$  is given by the following  expression
\widetext
\begin{eqnarray}
H\psi_{nk}&=&\frac{2\pi}{\hbar}\sum_{n'}\sum_{k'}\sum_{\bbox{q}}
|g_{\bbox{q}}|^2f^0(\varepsilon_{nk})(1-f^0(\varepsilon_{n'k'}))
\delta_{k,k'+q_x} (\psi_{nk}-\psi_{n'k'}) \nonumber \\
& & \times
\{\delta(\varepsilon_{n'k'}-\varepsilon_{nk}+\hbar\omega_{{q}})
(1+N^0(\omega_{{q}})) +
\delta(\varepsilon_{n'k'}-\varepsilon_{nk}-\hbar\omega_{{q}})
N^0(\omega_{{q}})\}.
\label{Hpsi}
\end{eqnarray}
\narrowtext
Here $g_{\bbox{q}}$ is the electron-phonon coupling
parameter, which will be specified below,
while $N^0(\omega_{{q}})$ is the equilibrium
phonon distribution. For simplicity we have assumed
that the phonon frequencies only depend on the
magnitude of $\bbox{q}$. In
Eq.\ (\ref{Hpsi}) we have explicitly written the factor
$\delta_{k,k'+q_x}$
discussed in the appendix. Utilizing the relationship
\begin{equation}
f(\varepsilon)(1\!-\!f(\varepsilon\!\pm\!\hbar\omega_q))  =
 (f(\varepsilon)\!-\!f(\varepsilon\!\pm\!\hbar\omega_q))
(1\!+\!N^0(\pm\omega_q)),
\end{equation}
the collision integral (\ref{Hpsi}) can be written as
\begin{eqnarray}
H\psi_{nk} &=&\frac{2\pi}{\hbar} \!
\sum_{n'\bbox{q}\sigma} |g_{\bbox{q}}|^2
\frac{\left|f^0(\varepsilon_{nk})
-f^0(\varepsilon_{nk}+\sigma\hbar \omega_q)\right|}
{4\sinh^2(\hbar\omega_q/2 k_{\rm B}T)} \nonumber \\
&& \times(\psi_{nk}\!-\!\psi_{n'k-q_x})
\delta(\varepsilon_{n'k-q_x}
\!-\!\varepsilon_{nk}\!-\!\sigma\hbar\omega_q),
\label{Hpsi2}
\end{eqnarray}

with  $\sigma$ assuming the two values $+1$ and $-1$ corresponding
to phonon absorption and emission respectively.

Our calculations of $g_{\bbox{q}}$ are based on
the standard electron-phonon interaction Hamiltonian, which
disregarding umklapp processes  takes the form
\begin{equation} \label{Hgen} \hat{H}_{ep} =
\sqrt{ \frac{\hbar}{2 V \rho \, \omega_q} } \sum_{\lambda,\bbox{q}}
M_{\lambda}(\bbox{q}) \hat{\rho}(\bbox{q})
(\hat{a}_{\bbox{q}} + \hat{a}_{-\bbox{q}}^{\dagger}),
\end{equation} where $\rho$ is the ion mass density, $V$ the
normalization
volume, $\hat{\rho}(\bbox{q})$ the Fourier-component of
the electron density operator, $\hat{a}_{-\bbox{q}}^{\dagger}$ a
phonon creation operator, and $\lambda$ the polarization index. The
coupling function $M$ is given by
\begin{equation} \label{Mgen}
M_{\lambda}(\bbox{q}) =
- V(\bbox{q}) \, \bbox{q}\! \cdot \!\bbox{\xi}_{\lambda},
\end{equation} $V(\bbox{q})$ being the electron-ion potential and
$\bbox{\xi}_{\lambda}$ a unit vector describing the polarization
$\lambda$. For a detailed treatment of
electron-phonon coupling in semiconductors see {\it e.g.\/}
Ref.~\onlinecite{Mahan72}.
The values of the GaAs parameters used in our calculations are
listed in Table~\ref{GaAsConst}.

\subsection{Coupling to acoustical phonons}
In GaAs-heterostructures at low temperatures
the electron-phonon scattering is mainly due to the combined
piezoelectric coupling and the deformation-potential coupling
\cite{Price,Mendez,Walukiewicz}. Below
we briefly sketch how these couplings are derived.

The deformation potential coupling only involves
the longi\-tudi\-nal acousti\-cal
 (LA) pho\-nons. The coupling to the transverse
acoustical (TA) phonons is suppressed by the square
of the ratio between the speed of sound and the speed of light.
In the   long-wavelength limit
 the coupling function $M^{df}$ for the deformation-potential
 coupling is written as
\begin{equation} \label{Mdf} M^{df} = \Xi \, q,
\end{equation} where $\Xi$, known as the deformation potential,
is the zero-wave-vector limit of $V(\bbox{q})$.

The electron-ion potential $V$ for the piezoelectric coupling is
found from the basic piezo-electric equations~\cite{Mahan72}.
For GaAs (zinc-blende structure) the only non-vanishing independent
piezoelectric constant is $h_{14}$ (reduced notation), and the
coupling function $M^{pz}_{\lambda}$ in this case becomes
\begin{equation} \label{Mpz} M^{pz}_{\lambda} =
i 2 e h_{14} (\hat{q}_x \hat{q}_y \hat{\xi}_{\lambda,z} +
              \hat{q}_y \hat{q}_z \hat{\xi}_{\lambda,x} +
              \hat{q}_z \hat{q}_x \hat{\xi}_{\lambda,y} ),
\end{equation} where $\hat{q_i} = (\bbox{q}/q)_i$ and
$\hat{\xi}_{\lambda,i} = (\bbox{\xi}_{\lambda})_i$. In the
piezoelectric case $\lambda$ is retained.

It is noted that $M^{df}$ is real while $M^{pz}$ is imaginary; thus
to second order the two terms do not interfere, and the
absolute square of the total coupling function
$M^{ac}$ is given by
\begin{equation} \label{Mac} |M^{ac}|^2 = |M^{df}|^2 + |M^{pz}|^2.
\end{equation}

To obtain a more tractable form of the piezoelectric coupling we
perform angular averages for the longitudinal and the (two)
transverse
modes separately and then add the terms\cite{Mahan72,Zook}.
 While this represents an approximation
compared to retaining the full ${\bf q}$-dependence of the coupling
 in the collision integral, we
expect it to involve only minor quantitative differences.
In adding the transverse and longitudinal contribution
we must remember the different average sound velocities
associated with each of the terms, originating in
the factor $1/\sqrt{\omega_q}=1/\sqrt{cq}$  in Eq.\ (\ref{Hgen}).
Expressing the transverse sound velocity as $x$ times the
longitudinal sound velocity we obtain
the following angular average of the
absolute square of the piezoelectric coupling function $M^{pz}$:
\begin{equation} \label{MpzAv}
|M^{pz}|^2 =
(eh_{14})^2 \left( \frac{12}{35} + \frac{1}{x} \frac{16}{35}
\right) \equiv P,
\end{equation} where we have introduced the constant $P$, and
where it is understood that the only sound velocity
appearing in the following is the longitudinal one.

The electron-phonon coupling parameter $g_{\bbox{q}}$ introduced in
Eq.~(\ref{Hpsi}) now becomes
\begin{equation} \label{gac}
|g_{\rm ac}(q)|^2 =
\frac{\hbar}{2 \rho V c}\: \frac{1}{q}(\Xi^2q^2+P) \:
|\langle nk|e^{i\bbox{q}\cdot\bbox{r}}| n'k'\rangle|^2.
\end{equation} The matrix element appearing in this expression is
treated in detail in the appendix.
 In writing Eq.\ (\ref{gac}) we have taken
the phonon frequency $\omega_q$ to be given by
\begin{equation}
\omega_q =cq.
\end{equation}

\subsection{Coupling to acoustical phonons
in the low-temperature limit}
In an analytical study of the low-temperature limit
we use the variational principle discussed in
section~3 above and choose a  trial function given by
\begin{equation}
U= U_{nk}={\rm sgn}(k).
\end{equation}
Thus $U$ is 1 on the branches corresponding to $k$ being positive,
while it is $-1 $ on the branches
corresponding to $k$ being negative.
This choice will lead to an expression for the conductivity which
in the case of a single occupied Landau-level and in the limit
of low temperature agrees with a numerical solution of the integral
equation. This suggests that the trial function is in fact the
exact one under these conditions.

First we evaluate the scalar product $(U,X)$ appearing in
the general expression (\ref{bound}) for the lower
bound on the conductivity.
By changing the integration variable to the
energy $\varepsilon_{nk}$ we get
\begin{equation}
(X,U)=\frac{eE}{\hbar\pi}\sum_nf^0(\varepsilon_{n0}),
\end{equation}
where $\varepsilon_{n0}$ is the $k=0$ value of $\varepsilon_{nk}$.
Because of the symmetry of the integral operator
$H$ given in Eq.~(\ref{Hpsi2}) the denominator occurring
in Eq.\ (\ref{bound}) becomes
\begin{eqnarray}
(U,HU)&=& \sum_{n}
\int_{-\infty}^{\infty}\frac{dk}{2 \pi}
\int_{-\infty}^{\infty}\frac{dk'}{2 \pi}
\sum_{n'} K_{nn'}(k,k') \nonumber \\
&& \times \frac{1}{2}
\left[1-{\rm sgn}(k) {\rm sgn}(k')\right]^2.
\label{varex}
\end{eqnarray}
As discussed in the appendix $k'=k-q_x$.
The occurrence of the factor $[1-{\rm sgn} (k) {\rm sgn}(k-q_x)]$
in the integrand of Eq.\ (\ref{varex})
implies that the summation over $q_x$ becomes restricted
by $q_x>k$ for $k>0$, while
$q_x <k$ for $k<0$. Using the symmetry with respect to
reversal of all momenta variables we may thus
restrict the $k$-integral to the interval 0 to $\infty$,
which limits $q_x$ to the region $q_x>k$.
If $\theta$ denotes the angle
between $\bbox{q}$ and the $x$-axis,
the integration over $\theta $ is therefore restricted by
\begin{equation}
\cos\theta > \frac{k}{q},\;\;\;\;0<q<\infty.
\end{equation}
Thus  we have to carry out
three integrations (over $q$, $k$ and $\cos\theta$)
as well as  two sums (over $n$ and $n'$).

Let us consider the simplest case, in which $n=n'=0$.
If the sound velocity is much less than the Fermi-velocity
then the energy-conserving delta-functions are
\begin{equation}
\delta(\varepsilon_{n'k'}\!-\!\varepsilon_{nk}\!\pm\!\hbar\omega_{{q}})
\simeq
\delta(\frac{\hbar^2}{2m}
(q^2\!\cos^2\theta\!-\!2kq\cos\theta)).\!
\end{equation}

The $q$-integral is now restricted to the interval $q>2k$,
while $k/q<\cos\theta<1$. This yields, after performing the
$\phi$-integral as shown in  the appendix,
\begin{eqnarray}
(U,HU)&=&
\frac{1}{4\pi^3\hbar^2c\rho}
\int_0^{\infty}
 \frac{d\varepsilon}{ v(\varepsilon)^2}
\int_{2k}^{\infty}dq \;
(\Xi^2q^2+P) \nonumber \\
&&\times\!I_{0,0}(q,2 \frac{k}{q})
\frac{\left|f^0(\varepsilon_{nk})
\!-\!f^0(\varepsilon_{nk}\!+\!\sigma\hbar \omega_q)\right|}{
4\sinh^2\left(\hbar\omega_q/2 k_{\rm B}T\right)},
\end{eqnarray}
where $I_{0,0}(q,2k/q)$ is given by Eq.~(A.5).
We may carry out the final integration over energy by expanding the
difference of the Fermi-functions in powers of $\omega_q$. Since the
contribution due to terms
involving higher order derivatives of the Fermi function
is seen to vanish by the use of partial integration, we obtain
\begin{eqnarray}
(U,HU)&=&\frac{1}{2\pi^3\hbar^2c\rho}\int_0^{\infty}
\frac{d\varepsilon}{v^2(\varepsilon)}
(-\frac{\partial f^0}{\partial \varepsilon}) \nonumber \\
&&\times \int_{2k}^{\infty}dq
\frac{(\Xi^2q^2+P)I_{0,0}(q,2k/q)\hbar\omega_q}
{4\sinh^2\left(\hbar\omega_q/2 k_{\rm B} T\right)}.
\end{eqnarray}
The remaining steps are standard.
 We assume that the temperature is much less than the Fermi
temperature, so that the integral over $\varepsilon$
yields $1/v_{\rm F}^2$, while $k$ may be replaced by
the Fermi momentum $k_{\rm F}$.
Furthermore, we assume that the temperature is small compared
to the characteristic temperature $\Theta$ defined by
\begin{equation}
k_{\rm B} \Theta =\hbar k_{\rm F}c.
\label{kTheta}
\end{equation}
Then we obtain
\begin{eqnarray}
(U,HU) &=&\frac{8}{\pi^2}\frac{m^2k_{\rm F}k_{\rm B} T}
{\rho c\hbar^4} \left( \Xi^2 + \frac{P}{4 k^2_{\rm F}} \right)
\nonumber \\
&& \times e^{-2k_{\rm F}^2\ell_h^2/(1+\gamma)}e^{-2\Theta/T}.
\end{eqnarray}
This results in the final conductivity expression
\begin{equation}
\sigma = \frac{2e^2}{h}l_{\sigma},
\label{sigmaac}
\end{equation}
where the  length  $l_{\sigma}$ is given by
\begin{equation}
l_{\sigma}=\frac{\pi c\hbar^3 \rho k_{\rm F}}{m^2
(4k_{\rm F}^2 \Xi^2 + P) }
e^{2k_{\rm F}^2\ell_h^2/(1+\gamma)} e^{2\Theta/T}.
\end{equation}
At low temperatures the conductivity thus increases exponentially,
in agreement with the result of the numerical
calculation (see Fig.\ 3).
In the presence of impurities, as we shall see in the following
subsection, the scattering against phonons
yields a contribution
to the inverse conductivity which is proportional to $T^3$
rather than $\exp(-2\Theta/T)$,
provided the temperature  is sufficiently low
 that the impurities dominate the scattering.

\subsection{Scattering from acoustical phonons and impurities}
When the impurity scattering dominates we may use
the variational principle with a trial
function which is proportional to the solution for
impurity scattering alone. We shall consider the case where
only the lowest Landau level, $n=0$, is
important. The trial function is thus chosen to be
proportional to $1/B_{0+}^{0-}$ as given by Eq.\ (\ref{ele}),
\begin{equation}
U(k)=k^2 \exp\left(2k^2l_h^2/[1+\gamma]\right).
\end{equation}
Since we have chosen the exact solution to the
impurity problem as our trial function, the calculated upper bound
on the  contribution from electron-phonon scattering,
$1/\sigma_{\rm ph}$, is exact, to lowest order in the magnitude
of the electron-phonon coupling. At low temperatures we may make the
approximation
\begin{equation}
(U(k)-U(k'))^2\simeq \left( \frac{dU(k)}{dk} \right)^2(k-k')^2,
\end{equation}
which is justified, since $k-k'=q_x$ and the restriction
to low temperatures implies that the phonon momentum is small.
Furthermore, we may neglect the deformation-potential coupling since
the piezoelectric coupling dominates for small $q$ according
to Eq.\ (\ref{gac}), and also set the matrix element
appearing in Eq.\ (\ref{gac}) equal to unity.
By inserting $U$ in Eq.\ (\ref{bound}) and carrying out the integrals
 we obtain
\begin{equation}
\sigma_{\rm ph}^{-1}=\frac{h}{2e^2}l_{\rm ph}^{-1}
\frac{3 \zeta(3)}{8\pi}
\frac{1+\gamma}{\gamma\tilde{\varepsilon}_{\rm F}^3}
\left(1+\frac{4\tilde{\varepsilon}_{\rm F}}{\gamma}\right)^2
\left(\frac{k_{\rm B}T}{\hbar\omega_h}\right)^3,
\end{equation}
where
$\tilde{\varepsilon}_{\rm F}={\varepsilon}_{\rm F}/\hbar\omega_h$,
and
\begin{equation}
l_{\rm ph}^{-1}=\frac{Pm^*}{\rho c^2\hbar^2}.
\end{equation}
This $T^3$-behavior agrees well with our numerical calculations
in the parameter range where it is expected to apply.
In Figs.\ 3 and 4 we show the
calculated conductivity, obtained by numerical solution of the
integral equation, for  samples with different
amounts of impurities.

\section{Scattering from optical phonons}
Next we investigate the combined effects of  scattering
by longitudinal optical phonons \cite{Mahan72}
and  impurities. The chief differences
from the previous section involve the $q$-dependence of
the electron-phonon matrix-element and the absence of dispersion
in the phonon frequencies. The conductivity is obtained by
solving the Boltzmann equation in the limit where the temperature
is much less than $\hbar\omega_0/k_{\rm B}$.

\subsection{Coupling to optical phonons}
We shall use the simple model where the phonon
frequency is independent of momentum,
\begin{equation}\label{omegaopt}
\omega_q \simeq \omega_0.
\end{equation}
The electron-phonon matrix element~\cite{Mahan72} is given by
\begin{equation}
|g(nn',\bbox{q})|^2 = \frac{1}{q^2}\frac{M_0^2}{V} M_n^{n'}(u),
\end{equation}
where $V$ is the normalization volume and where
\begin{equation}
M_0^2=2\pi\frac{e^2\hbar \omega_0}{\varepsilon_0}
\left(\frac{1}{\kappa_\infty}-\frac{1}{\kappa_0}\right).
\end{equation} The function $M$ is given by
\begin{equation}
M_n^{n'}(u) =
\left|\langle nk|e^{i \bbox{q}\cdot\bbox{r}}|n'k'\rangle\right|^2,
\end{equation} which is calculated in the appendix,
where also $u$ is defined.

The approximation in Eq.\ (\ref{omegaopt}) allows us to integrate
over the $y$ and $z$ components of the phonon momentum. Using a
delta-function for the wave-function in the $z$-direction, we get
\begin{equation}
\sum_{q_yq_z} |g(nn',\bbox{q})|^2 =
\frac{M_0^2}{(2\pi)^2L} K_n^{n'}(q_x),
\end{equation}
where
\begin{equation} \label{Knnqx}
K_n^{n'}(q_x) = \int_{-\infty}^\infty
\frac{dq_y}{\sqrt{q_x^2+q_y^2}} M_n^{n'}(u(q_x,q_y)).
\end{equation}

\subsection{Scattering from optical phonons and impurities}
We now consider the case of impurity and optical phonon
scattering. The Boltzmann equation in this  case reads
\begin{equation}
\frac{eE}{k_{\rm B}T} v_{nk} f^0_{nk}(1-f^0_{nk})
=\left( \frac{\partial f_{nk}}{\partial t}\right)_{\rm coll}^{\rm imp}
+\left( \frac{\partial f_{nk}}{\partial t}\right)_{\rm coll}^{\rm op}.
\end{equation}
The collision integral for the phonon scattering
given in Eq.\ (\ref{Hpsi2}) simplifies in the case of
optical phonons to
\widetext
\begin{eqnarray}
\left( \frac{\partial f_{nk}}{\partial t}\right)_{\rm coll}^{\rm op}
&=& -\frac{M_0^2}{\hbar 2\pi L}
\frac{1}{4\sinh^{2}(\hbar\omega_0/2 k_{\rm B} T)}
\sum_{n'q_x}\sum_\sigma K_n^{n'}(q_x) \left|f^0(\varepsilon_{nk})
-f^0(\varepsilon_{nk}+\sigma\hbar \omega_0)\right|\nonumber\\
&&
\times \left(\psi_{nk}-\psi_{n'k-q_x}\right)
\delta(\varepsilon_{n'k-q_x}-\varepsilon_{nk}-\sigma\hbar \omega_0),
\end{eqnarray}
\narrowtext
with $K_n^{n'}(q_x)$ given in Eq.\ (\ref{Knnqx}).
The argument of the energy-conserving delta-function is zero for
\begin{equation}
q_x =  k \left(1-s\sqrt{1+\frac{\hbar\omega_h(n-n')
+\sigma \hbar\omega_0}{\hbar^2 k^2/2m}}\right),
\end{equation}
for each value of $nk$, provided  the square root is real.
The branch  index $s$ assumes the values $+1$ and $-1$ corresponding
to forward and backward scattering.
The Boltzmann equation now simplifies to
\begin{eqnarray}
1&=&\sum_{n'\;s'}B_{n+}^{n's'}(\varepsilon)
(\phi_{n}(\varepsilon)-s'\phi_{n'}(\varepsilon)) \nonumber \\
&& + \sum_{n'\;s'}C_{n+}^{n's'}(\varepsilon,\sigma)
(\phi_{n}(\varepsilon)- s'\phi_{n'}
(\varepsilon+\sigma\hbar\omega_0)),
\label{matr1}
\end{eqnarray}
where the matrix $B$ is given by Eq.\ (\ref{ele}).
The optical-phonon scattering gives rise to
the matrix $C$, which is given by
\widetext
\begin{eqnarray}
C_{ns}^{n's'}(\varepsilon,\sigma) &= &
\frac{1}{f^0(\varepsilon)[1-f^0(\varepsilon)]}
\frac{M_0^2}{\hbar^2(2\pi)^2}
\frac{1}{4\sinh^{2}(\hbar\omega_0/2 k_{\rm B} T)}
\nonumber \\
&& \times \sum_{\sigma } \left|f^0(\varepsilon)-f^0(\varepsilon+
\sigma\hbar \omega_0)\right|\frac{K_{n}^{n'}(sk_n(\varepsilon)-s'
k_{n'}(\varepsilon+\sigma\hbar\omega_0))}
{v_n(\varepsilon)v_{n'}(\varepsilon+\sigma\hbar\omega_0)}.
\end{eqnarray}
\narrowtext
In GaAs the typical optical phonon energies are about 36 meV.
Furthermore, since the cyclotron energy  $\hbar \omega_c$ in GaAs
is $1.5$ meV $B$/T, the
typical situation will be that the energy spacing between
the Landau levels as modified by the confinement
potential, $\hbar\omega_h$, is much
smaller than the phonon energies, $\hbar\omega_0 \gg \hbar\omega_h$.
Since we are interested in temperatures that are small
or comparable in magnitude to the energy spacing $\hbar\omega_h$,
 we are therefore always in a situation
where $k_{\rm B}T$ is much less than
$\hbar\omega_0$. Under these circumstances it
is possible to simplify
the Boltzmann equation for electrical transport by considering
it to be a coupled system of equations for the
functions $\phi(\varepsilon\pm n\hbar\omega_0) $ with $n$ integer.
Since the current is mainly determined
by the distribution of electrons $\phi(\varepsilon)$
within a thermal layer
of thickness comparable to $k_{\rm B}T$, it is sufficiently accurate
to neglect the contributions from
$\phi(\varepsilon \pm n\hbar\omega_0) $,
when $n$ is different from zero. The
collision integral may therefore be approximated by
 the ``scattering-out'' term, corresponding to the neglect
of vertex-corrections in the Kubo-approach.

In calculating the conductivity from Eq.\ (\ref{matr1})
we  thus neglect the ``scattering-in'' terms involving
$\phi_{n'}(\varepsilon+\sigma\hbar\omega_0)$, while
 retaining the full collision matrix for the impurities.
We have then evaluated the solution of Eq.\ (\ref{matr1})
numerically and inserted the solution in
the conductivity formula (\ref{cond}). Results are shown in
Fig.\ 5. The low-temperature behavior is dominated
by the impurities, which give rise to an initial increase
of the conductivity, in accordance with Eq.\ (\ref{cond1}).
The optical phonons
come in at higher temperatures, yielding a maximum of
the conductivity as a function of temperature. The position of the
maximum is roughly proportional to the
logarithm of the  strength of the impurity scattering,
since the phonon contribution depends exponentially on temperature.
Note however that the acoustical phonons have
reduced the conductivity for the high mobility cases rather strongly
at the temperature range shown here, see Fig.\ 4.
The dashed line represents the conductivity without optical
phonon scattering.

In comparing the  curves Figs.\ 3-5 we observe
that the optical phonon scattering in  clean systems
tends to dominate
the total scattering already at fairly low temperatures,
around 50~K. This contrasts the situation
in typical GaAs-based two-dimensional electron gases\cite{Walukiewicz},
where  the optical phonon scattering begins to
dominate around 100~K.

\section{Conclusion}
We have shown that the magnetoconductivity of quantum wires may
be discussed in a simple and unified fashion within the framework
of a Boltzmann equation, by taking into account
the influence of the magnetic field on the electron group-velocity
and the matrix-elements governing the
transition probability. By treating in detail
the scattering from acoustical as well as optical phonons
in the presence of impurity scattering, we have
determined the temperature dependence of
the magnetoconductivity for realistic choices of parameters
in GaAs-based structures.
In particular, we have found that the magnetoconductivity
exhibits a maximum as a function of temperature,
 depending on the relative
strength of the  impurity and electron-phonon scattering. The
calculated  magnetoconductivity
 oscillates when the Fermi energy or the magnetic field is varied.
Our detailed calculations show that the scattering
against optical phonons  in quantum wires
is significant at  temperatures somewhat smaller
than the corresponding temperatures for the two-dimensional case.

In order to relate our theoretical results to
experiment \cite{fowler,kastner,motta}
it is necessary to consider not only the diagonal conductivity
element $\sigma_{xx}$, but also $\sigma_{xy}$.
To compare with experiment one must therefore
either determine $\sigma_{xy}$ theoretically or
compare our calculated $\sigma_{xx}$ to simultaneous
measurements of $\rho_{xx}$ and $\rho_{xy}$. Alternatively,
 our  results could be compared to experiments that use a
Corbino-type geometry. The  effects
 predicted in this paper should be observable in quantum wires of
sufficient purity, since otherwise
the electron density may vary considerably  due to fluctuations in
the electrostatic potential from the donors.

\section*{Acknowledgements}
We want to thank A.D.~Stone and R.G.~Wheeler for helpful discussions.
H.B.\ is supported by Grant No.~11-9454 from the Danish Natural Science
Research Council.

\appendix
\section*{}
In this appendix we study the square of the matrix element
$\langle nk|\exp(i \bbox{q}\! \cdot \! \bbox{r})|n'k' \rangle$.
Using the single electron wave functions in Eq.\ (\ref{eigen}) and
performing the $x$- and $y$-integrals one obtains
\begin{eqnarray}
|\langle nk|\exp(i \bbox{q} \!\cdot\! \bbox{r})|n'k' \rangle|^2 &=&
\delta_{k,k'+q_x} \,
\frac{n_{\rm min}!}{n_{\rm max}!}
u^{|n-n'|} \nonumber \\
&& \times \left[
L_{n_{\rm min}}^{|n-n'|}(u) \right]^2\, e^{-u},
\label{eiqr}
\end{eqnarray} where $n_{\rm min} = {\rm min}(n,n')$ and
$n_{\rm max} = {\rm max}(n,n')$; $L^m_n(u)$
are the Laguerre polynomials
while $u$ is given by
\begin{equation} \label{u}
u= \frac{1}{1+\gamma}\frac{1}{2} (q_x\ell_h)^2 +
\frac{1}{2} (q_y\ell_h)^2.
\end{equation} The Kronecker delta appearing in Eq.\ (\ref{eiqr}),
and which is written explicitly in the expression for the
integral operator $H$ in Eq.\ (\ref{Hpsi}), is a
consequence of the translational invariance along the
$x$-direction.

In calculations involving acoustical phonons it is natural to use the
polar coordinates $(q,\theta,\phi)$ with the $x$-axis as the polar axis
so that $q_x = q \cos\theta$, $q_y = q \sin\theta \cos\phi$, and
$q_z = q \sin\theta \sin\phi$. According to Eqs.\ (\ref{Hpsi})
and (\ref{gac}) the only $\phi$-dependence is through $u$ defined
above. It is therefore of interest to calculate the integral
\begin{equation} \label{Inn}
I_{n,n'}(q,\cos\theta) \equiv \int_0^{2 \pi} d\phi \, u^{|n-n'|}
\left[ L_{n_{\rm min}}^{|n-n'|} \right]^2 \, e^{-u}.
\end{equation} From Eq.\ (\ref{u}) it follows that
the $\phi$-dependence of $u$
is of the form $u = \beta + \alpha \cos^2\phi$, so the integrand of
$I_{n,n'}$ therefore takes the form of a polynomial in $\cos^2\phi$
times $\exp(-\alpha \cos^2\phi)$.
The integral over $\phi$ may then be carried out explicitly in terms
of the beta-function and  Gauss' hypergeometric function.

In the case where we consider only the lowest Landau level we just
need to calculate $I_{0,0}$. Using the results quoted above we find
\begin{eqnarray}
I_{0,0}(q,q_x/q) &=& 2 \pi
\exp\left(\!-\frac{1}{2}(q_x\ell_h)^2
\frac{1}{1\!+\!\gamma}\!-\!\frac{1}{4} \ell_h^2(q^2 \!- \!q_x^2)\!\right),
\nonumber \\
&& \times {\rm I}_0
\left( \frac{1}{4} \ell_h^2(q^2 - q_x^2) \right)
\label{I00}
\end{eqnarray}
where I$_0(x)$ is the 0$^{th}$ Bessel function of imaginary argument.

\narrowtext
\begin{table}
\caption{The GaAs constants used in this paper. Unless otherwise
indicated the values are taken from
Ref.~\protect\onlinecite{Blakemore}.}
\begin{tabular}{l@{\hspace{0mm}}c@{\hspace{0mm}}r@{\hspace{0mm}}l}
parameter & symbol & \multicolumn{2}{c}{value} \\ \hline
ion mass density &
$\rho$ & 5.3$\times$10$^{3}$ & $\,$kg m$^{-3}$ \\
longitudinal sound velocity &
$c$ & 5.2$\times$10$^{3}$ & $\,$m s$^{-1}$ \\
transverse sound velocity &
$xc$ & 3.0$\times$10$^{3}$ & $\,$m s$^{-1}$ \\
sound velocity ratio  & $x$ & 0.58 & \\
static dielectric constant & $\kappa_0$ & 12.8 &\\
high-frequency dielectric constant & $\kappa_\infty$ & 10.6 & \\
piezoelectric constant \protect\cite{Zook}&
$h_{14}$ & 1.38$\times$10$^{9}$ & $\,$V m$^{-1}$\\
piezoelectric coupling, Eq.~(\protect\ref{MpzAv})
& $P$ & 5.4$\times$10$^{-20}$  & $\,$J$^2$ m$^{-2}$ \\
deformation potential  \protect\cite{Mendez}
& $\Xi$ & 2.2$\times$10$^{-18}$ & $\,$J\\
effective electron mass & $m^*$ & 0.067 & $\,$$m_0$\\
optical phonon energy & $\hbar\omega_0$ & 36 & $\,$meV \\
\end{tabular}
\label{GaAsConst}
\end{table}

\begin{figure}
\caption{The two lowest $k$-dependent
energy bands $\varepsilon_{0k}$ and $\varepsilon_{1k}$
of the quantum wire. The dashed line represents an arbitrary
Fermi level above the bottom of the second band.}
\label{fig1}
\end{figure}

\begin{figure}
\caption{The conductivity for a quantum wire with
impurity scattering plotted versus the
Fermi level for three different choices of the confinement
potential corresponding to $\gamma=$ 1, 0.5, and 0.25.
The dashed line is the $T=0$ result and the solid lines correspond
to ${\rm k_{{\rm B}}} T=0.05 \; \hbar \omega_h$.
The magnetic field is 9 T.}
\label{fig2}
\end{figure}

\begin{figure}
\caption{Plots of the normalized conductivity versus temperature
for a GaAs quantum wire at the magnetic field $B=9$~T,
the confinement parameter
$\gamma=0.5$, and the Fermi level
$\varepsilon_{\rm F}=0.6$~$\hbar \omega_h$.
Taking into account both impurity scattering,
Eq.~(\protect\ref{matr}),  and acoustical phonon scattering,
Eqs.~(\protect\ref{Hpsi2}) and~(\protect\ref{gac}),
the five dotted curves, $\sigma_1,\ldots,\sigma_5$, are numerical
results for the zero-field, zero-temperature mobilities,
$\mu_{\rm imp} = $ 7.5, 75,
300, 3000, and 30000~m$^2$/(Vs) respectively.
  The last two rather unrealistic
high mobilities are considered to allow a study of the transition
from impurity-dominated scattering to phonon-dominated scattering
at temperatures low enough ($T <\Theta \simeq 4.0 $ K -- see
Eq.~(\protect\ref{kTheta})) for the approximative
result Eq.~(\protect\ref{sigmaac}) to apply.
The two full curves, $\sigma_{a4}$ and $\sigma_{a5}$,
are plots of the conductivity
(rescaled to match $\sigma_4$ and $\sigma_5$)
calculated from Eq.~(\protect\ref{sigmaac})
where only the acoustical phonon scattering is present.
The full curve $\sigma_{\rm imp}$ is the case
where only impurity scattering is present.
The two full curves, $\sigma_{t4}$ and $\sigma_{t5}$, are the
results of assuming that the inverse conductivities for each
scattering mechanism considered separately may be added,
$\sigma_{tj} = \sigma_{aj} \sigma_{\rm imp}/(\sigma_{aj} +
\sigma_{\rm imp})$ ($j=4,5$),
to approximate the exact numerical calculations of the total
conductivity.}
\label{fig3}
\end{figure}

\begin{figure}
\caption{Plots of the normalized conductivity versus temperature
for a GaAs quantum wire at the magnetic field $B=9$~T and
the Fermi level $\varepsilon_{\rm F} = 0.6$~$\hbar \omega_h$.
The confinement parameter in panel (a) is $\gamma=1.0$
and in (b) it is $\gamma=0.5$.
The dashed curve in each panel is the case where only impurity
scattering is taken into account. The four full curves in each
panel is the result of combining
impurity and acoustical phonon scattering  for each of
the following values of the
 zero-field, zero-temperature mobility, $\mu_{\rm imp}= $ 0.9, 9, 90,
and 900~m$^2$/(Vs).}
\label{fig4}
\end{figure}

\begin{figure}
\caption{The conductivity with combined impurity and
optical phonon scattering for a GaAs quantum wire.
The normalized  conductivity is plotted versus temperature
for different choices of confinement potential strengths, Fermi
energies and mobilities. The magnetic field is $B=9$~T.
The confinement parameter is
$\gamma=1$ for panel (a) and (c) and $\gamma=0.5$ for (b) and (d);
the Fermi level is $\varepsilon_{\rm F} = 0.6 \; \hbar\omega_h$,
{\it i.e.}, close to
the bottom of the first band, for (a) and (b), and panel (c) and (d)
have $\varepsilon_{\rm F}=1.3 \; \hbar\omega_h$. The conductivity is
shown for each of the following values of the zero-field,
zero-temperature mobility, $\mu_{\rm imp}= $ 0.9, 9, 90, and
900~m$^2$/(Vs).
The dashed curve in each panel is the case where only impurity
scattering is taken into account.}
\label{fig5}
\end{figure}


\begin{thebibliography}{99}

\bibitem{beenakker} For a recent review, see
C.\ W.\ J.\ Beenakker and H.\ van Houten,\\
Solid State Physics {\bf 44}, 1  (1991).

\bibitem{martin} T.~Martin and S.~Feng,
Phys.~Rev.~Lett.\ {\bf 64}, 1971 (1990),
Phys.~Rev.~B {\bf 44}, 9084 (1991).

\bibitem{maslov}
D.\ L.\ Maslov, Y.\ B.\ Levinson, and S.\ M.\ Badalian,
Phys.~Rev.~B {\bf 46}, 7002 (1992).

\bibitem{komi}
S.\ Komiyama, H.\ Hirai, M.\ Ohsawa, Y.\ Matsuda, S.\ Sasa,
and T.\ Fujii, Phys.~Rev.~B {\bf 45}, 11085 (1992).

\bibitem{ando}  H.\ Akera and T.\ Ando,
Phys.\ Rev.\ B {\bf 41}, 11967 (1990).

\bibitem{Kubo} R.\ Kubo,
J.\ Phys.\ Soc.\  of Japan {\bf 12}, 570 (1957).

\bibitem{vasilo}
P.\ Vasilopoulos, P.\ Warmenbol, F.\ M.\ Peeters,
and J.\ T.\ Devreese,\\
 Phys.~Rev.~B {\bf 40}, 1810 (1989).

\bibitem{mori} N.\ Mori,  H.\ Momose, and C.\ Hamaguchi,
Phys.\ Rev.\ B {\bf 45}, 4536 (1992).

\bibitem{bruus} H.\ Bruus and K.\ Flensberg,
J.~Phys.:~Condens.~Matter {\bf 4}, 9131 (1992).

\bibitem{tso}      H.\ C.\ Tso and
P.\ Vasilopoulos, Phys.\ Rev.\ B {\bf 44}, 12952 (1991).

\bibitem{boiko}
I.\ I.\ Boiko, P.\ Vasilopoulos and V.\ I.\ Sheka,
Phys.~Rev.~B {\bf 46}, 7794 (1992).

\bibitem{wagner}
M.\ Wagner, Phys.\ Rev.\ B {\bf 45}, 11606 (1992)

\bibitem{Landauer} R.\ Landauer, Phil.\ Mag.\ {\bf 21}, 768 (1970).

\bibitem{buttiker} M.\ B\"{u}ttiker, Phys.\ Rev.\ Lett.\ {\bf 57},
1761 (1986).

\bibitem{shepard}  K. Shepard, Phys.~Rev.~B {\bf 44}, 9088 (1991).

\bibitem{suhrke} M.\ Suhrke and S.\ Wilke,
Phys.\ Rev.\ B {\bf 46}, 2400 (1992).

\bibitem{staring}
A.\ A.\ M.\ Staring, H.\ van Houten, C.\ W.\ J.\ Beenakker, and
C.\ T.\  Foxon, Phys.~Rev.~B {\bf 45}, 9222 (1992).

\bibitem{transport} For a discussion of the
Boltzmann equation see for instance
Henrik Smith and H.~H\o jgaard Jensen, {\it Transport Phenomena},
Oxford University Press 1989.

\bibitem{Mahan72} G.D.~Mahan, {\em Polarons in Ionic Crystals and Polar
Semiconductors}, J.T.~Devreese (ed.),\\
North-Holland/American Elsevier, p.~553 (1972).

\bibitem{Price} P.J.~Price, Surf.~Sci. {\bf 113}, 119 (1982) and
{\bf  143}, 145 (1984).

\bibitem{Mendez} E.E.~Mendez, P.J.~Price, and M.~Heiblum,
Appl.~Phys.~Lett. {\bf 45}, 294 (1984).

\bibitem{Walukiewicz} W.~Walukiewicz, H.E.~Ruda, J.~Lagowski,
and H.C.~Gatos, Phys.~Rev.~B {\bf 29}, 4818 (1984).

\bibitem{Zook} J.D.~Zook, Phys.~Rev.~A {\bf 136}, 869 (1964).

\bibitem{fowler}
A.\ B.\ Fowler, A.\ Hartstein, and R.\ A.\ Webb,
Physica {\bf 117-118B}, 661 (1983).

\bibitem{kastner}
M.\ A.\ Kastner, S.\ B.\ Field, J.\ C.\ Licini, and S.\ L.\ Park,
Phys.~Rev.~Lett.\ {\bf 60}, 2535 (1988).


\bibitem{motta}
R.\ Mottahedeh, M.\ Pepper, R.\ Newbury, J.\ A.\ A.\ J.\ Perenboom,
and K.-F.\ Berggren, Solid State Comm.\ {\bf 72}, 1065 (1989).

\bibitem{Blakemore} J.S.~Blakemore, J.Appl.Phys. {\bf 53}, R123 (1982)

\end{thebibliography}
\end{document}